\documentstyle [12pt,epsf] {article}

\newcommand{\sect}[1]{ \section{#1} \setcounter{equation}{0} }
\newcommand{\subsect}{\subsection}

\newcommand{\req}[1]{(\ref{#1})}

\newcommand{\nwc}{\newcommand}
\nwc{\btu}{\bigtriangleup}
\nwc{\cd}{\cdot}
\nwc{\zd}{{\bf Z}$_3$\ }

\nwc{\hyp} {\hyphenation}
\hyp{orbi-fold} 
\hyp{theo-ries}
\hyp{theo-ry}
\hyp{regu-lari-zation}

\newcommand{\Z}{\ZZ}
\def\bfone{\relax{\rm 1\kern-.35em 1}}
\def\inbar{\vrule height1.5ex width.4pt depth0pt}
\def\IC{\relax\,\hbox{$\inbar\kern-.3em{\mss C}$}}
\def\ID{\relax{\rm I\kern-.18em D}}
\def\IF{\relax{\rm I\kern-.18em F}}
\def\IH{\relax{\rm I\kern-.18em H}}
\def\II{\relax{\rm I\kern-.17em I}}
\def\IN{\relax{\rm I\kern-.18em N}}
\def\IP{\relax{\rm I\kern-.18em P}}
\def\IQ{\relax\,\hbox{$\inbar\kern-.3em{\rm Q}$}}
\def\IR{\relax{\rm I\kern-.18em R}}
\def\ZZ{\relax{\hbox{\mss Z\kern-.42em Z}}}
\font\cmss=cmss10 \font\cmsss=cmss10 at 7pt
\def\ZZ{\relax\ifmmode\mathchoice
{\hbox{\cmss Z\kern-.4em Z}}{\hbox{\cmss Z\kern-.4em Z}}
{\lower.9pt\hbox{\cmsss Z\kern-.4em Z}}
{\lower1.2pt\hbox{\cmsss Z\kern-.4em Z}}\else{\cmss Z\kern-.4em
Z}\fi}

\nwc{\E}{\hat{E}}
\nwc{\ten}{ten--dimensional}
\nwc{\four}{four--dimensional}
\nwc{\gev} {{\rm GeV}}
\nwc{\tev} {{\rm TeV}}
\nwc{\mP} {$M_{\rm Planck}$}
\nwc{\mx} {$M_{\rm X}$}
\nwc{\ms} {$M_{\rm string}$}
\nwc{\sieb}{\mbox{\boldmath $\ov{27}$}}
\nwc{\sie}{\mbox{\boldmath ${27}$}}

\nwc{\be}  {\begin{equation}}
\nwc{\ee}  {\end{equation}}
\nwc{\ba}  {\begin{array}}
\nwc{\ea}  {\end{array}}
\nwc{\bdm} {\begin{displaymath}}
\nwc{\edm} {\end{displaymath}}

\nwc{\bea} {\be\ba{lcl}}
\nwc{\eea} {\ea\ee}

\nwc{\bda} {\bdm\ba{lcl}} 
\nwc{\eda} {\ea\edm}

\nwc{\bc}  {\begin{center}}
\nwc{\ec}  {\end{center}}

\nwc{\ds}  {\displaystyle}
\nwc{\bmat}{\left(\ba}
\nwc{\emat}{\ea\right)}
\nwc{\nn} {\nonumber}
\nwc{\nnn} {\nonumber \vspace{.2cm} \\ }
\nwc{\ra}{\rightarrow}
\nwc{\lra}{\longrightarrow}

\nwc{\p} {\partial}

\nwc{\scr}  {\scriptstyle}
\nwc{\tx}  {\textstyle}
\nwc{\scs} {\scriptscriptstyle}

\nwc{\ov}  {\overline}

\nwc{\hb}  {\bar h}
\nwc{\xb}  {\bar x}
\nwc{\yb}  {\bar y}
\nwc{\zb}  {\bar z}
\nwc{\wb}  {\bar w}
\nwc{\Ob}  {\bar O}
\nwc{\Yb}  {\bar Y}

\nwc{\ep} {\epsilon}
\nwc{\de} {\delta}
\nwc{\Th} {\Theta}
\nwc{\th} {\theta}
\nwc{\al} {\alpha}
\nwc{\si} {\sigma}
\nwc{\Si} {\Sigma}
\nwc{\om} {\omega}
\nwc{\Om} {\Omega}
\nwc{\Ga} {\Gamma}
\nwc{\ga} {\gamma}
\nwc{\bet} {\beta}
\nwc{\La} {\Lambda}
\nwc{\la} {\lambda}

\nwc{\Sc}  {{\cal S}}
\nwc{\Rc}  {{\cal R}}
\nwc{\Dc}  {{\cal D}}
\nwc{\Oc}  {{\cal O}}
\nwc{\Cc}  {{\cal C}}
\nwc{\gc}  {{\cal g}}

\nwc{\Of}  {{\cal O}_f}
\nwc{\Oft} {{\cal O}_{f_2}}
\nwc{\Ofo} {{\cal O}_{f_1}}

\nwc{\Pc}  {{\cal P}}
\nwc{\Mc}  {{\cal M}}
\nwc{\Ec}  {{\cal E}}
\nwc{\Fc}  {{\cal F}}

\nwc{\Hc}  {{\cal H}}
\nwc{\Kc}  {{\cal K}}
\nwc{\Wc}  {{\cal W}}

\nwc{\Fcp} {{\cal F}^\pr}
\nwc{\Hcp} {{\cal H}^\pr}

\nwc{\Xc}  {{\cal X}}
\nwc{\Gc}  {{\cal G}}
\nwc{\Zc}  {{\cal Z}}
\nwc{\Nc}  {{\cal N}}

\nwc{\xc}  {{\cal x}}
\nwc{\Ac}  {{\cal A}}
\nwc{\Bc}  {{\cal B}}
\nwc{\Uc} {{\cal U}}
\nwc{\Vc} {{\cal V}}
\nwc{\Lc} {{\cal L}}
\nwc{\Qc} {{\cal Q}}

\nwc{\lng} {\langle}
\nwc{\rng} {\rangle}

\nwc{\lf} {\left}
\nwc{\ri} {\right}

\nwc{\diag} {{\rm diag}}
\nwc{\inv}  {{\rm inv}}
\nwc{\mod}  {{\ \rm mod\ }}
\nwc{\dete}  {{\rm det}}
\nwc{\tr}  {{\rm tr}}
\nwc{\im}  {{\rm Im}}
\nwc{\re}  {{\rm Re}}

\nwc{\h} {\frac{1}{2}}
\nwc{\fc} {\frac}

\def\KK{\relax{\rm I\kern-.18em K}}
\def\RR{\relax{\rm I\kern-.18em R}}
\def\NN{\relax{\rm I\kern-.18em N}}
\def\PP{\relax{\rm I\kern-.18em P}}

\def\zz{\relax{\sf Z\kern-.3em Z}}
\def\ZZ{\relax{\sf Z\kern-.4em Z}}
\def\ZZZ{{\relax{\sf Z}\kern -.5em Z}}
\def\ZZZ{Z\kern -0.37em Z}
\def\QQ{{\rm \kern .25em
             \vrule height1.4ex depth-.12ex width.06em\kern-.31em Q}}
\def\CC{{\rm \kern .25em
             \vrule height1.4ex depth-.12ex width.06em\kern-.31em C}}
\if@twoside\oddsidemargin25mm
\else\oddsidemargin25mm\evensidemargin25mm\marginparwidth25mm\fi%
\begin{document}
\begin{titlepage}
{\sf \begin{flushright}
{CERN--TH/97--18}\\
{NEIP--009/96}\\
{hep--th/9702110}\\
{February 1997}
\end{flushright}}
\vfill
\vspace{0cm}
\begin{center}
{\large\bf String--Unification, Universal One--Loop Corrections \\[4mm] 
and Strongly Coupled Heterotic String Theory$^{\mbox{\boldmath $\ast$}}$}
\end{center}

\vskip 1.2cm
\begin{center}
{\sc H.P. Nilles$^{1,\ast\ast}$}   {\ \ \small and\ \ }   
{\sc S. Stieberger$^2$}\\
\vskip 1cm
{\em $^1$CERN,  Theory Division}\\
{\em CH--1211 Gen\`eve, SWITZERLAND}\\
\vskip .5cm
{\small and}\\
\vskip .5cm
{\em $^2$Institut de Physique Th\'eorique}\\
{\em Universit\'e de Neuch\^atel}\\
{\em CH--2000 Neuch\^atel, SWITZERLAND}
\end{center}
\vfill

\thispagestyle{empty}
\vskip 2cm
\begin{abstract}
We derive the universal threshold corrections in heterotic string
theory including a continuous Wilson line. Unification of gauge and
gravitational couplings is shown to be possible even within perturbative
string theory.
The relative importance  of gauge group
dependent and independent thresholds on unification 
is clarified. Equipped with these results we can then attempt an
extrapolation to the strongly coupled heterotic string --- M--theory.
We argue that such an extrapolation might be meaningful because of
the holomorphic structure of the gauge coupling function and the
close connection of the threshold corrections to the anomaly
cancelation mechanism. 
\end{abstract}

\vskip 5mm \vskip0.5cm
\hrule width 5.cm \vskip 1.mm
{\small\small  $^\ast$ Supported by the
Swiss National Science Foundation, the European Commission TMR programmes 
ERBFMRX--CT96--0045 and ERBFMRX--CT96--0090  as well as a grant from
Deutsche Forschungsgemeinschaft SFB-375-95.\\
$^{\ast\ast}$ On leave from: Physik Department, Technische 
Universit\"at M\"unchen, D--85747 Garching, FRG.}
\end{titlepage}

\sect{Introduction}

The framework of string theory might ultimately give an explanation
of the unification of all fundamental coupling constants: gauge and
gravitational. At a very naive level such an unification is obtained,
although one could still be dissatisfied with the numerical
precision of this statement,  in the heterotic
theory the unification scale seems to be
a factor 20 smaller than
the string scale. Related to this question is the fact
that we do not really understand the reason why the string coupling
is so small to allow perturbative unification. Some aspects of unification
might even look more ``natural'' when viewed from a strong coupling
point of view \cite{newwitten}, 
although our ability to do explicit calculations is very
limited (if not nonexistent) in this region.

In that sense there is at the moment no alternative to threshold
calculations in {\it perturbative} string theory. One might hope that
such results could be extended to the region of stronger coupling.
A study of these questions will be presented in this
paper. By generalizing previous work, 
we shall first give the results of a full gauge coupling
threshold calculation in the presence of a continous Wilson line
background. Within this framework we show that unification can
be achieved in the perturbative theory with rather natural values 
of the moduli. This can be seen as a consequence of the so--called
gauge group dependent threshold corrections \cite{ns}. 
The new results of this 
paper concern also the gauge group independent thresholds which are
found less relevant for the mechanism of unification. They are,
however, very important for the understanding of the structure of
the holomorphic gauge kinetic function $f(\phi)$ and its extrapolation
to the region of stronger coupling.

There are reasons to believe that these threshold functions computed 
explicitly in perturbative heterotic string theory contain crucial
information about the full underlying string or M--theory. The first
reason is the intimate connection to the mechanism of anomaly
cancelation in heterotic string theory \cite{GS}. The existence of the
thresholds can be deduced from the presence of the Green--Schwarz terms
required by the anomalies \cite{DIN,in86}.
The second reason resides in the
holomorphic structure of the $f$--function that, due to 
nonrenormalization theorems \cite{SV,N}, is easily controllable in perturbation
theory. This could then imply that the results obtained here might be
of more general validity beyond the weak coupling regime.

In this paper we shall present the threshold calculation in the
presence of a continuous Wilson line in detail, discuss its relevance
for the question of unification and argue that some aspects of it
will carry over to the M--theory \cite{Mtheory,HoravaW}
domain. The paper will be organized as
follows. In section 2 we give for completeness a discussion of
the definition of the string coupling constant $g_{\rm string}$ at the
1-loop level while in section 3 we summarize the constraints for 
models that are consistent with the requirements of gauge coupling
unification. Section 4 contains the new results of the threshold
calculation with some technical details relegated to the appendix.
We comment especially on the universal one--loop corrections and its 
implication for unification. In section 5 we consider the possible
relevance of this calculation in the framework of M--theory. We shall
clarify the connection to the cancelation of anomalies and the possible
uses of the holomorphicity of the gauge kinetic function. We also study
the range of validity of the linearized approximation concerning the
$T$, $U$ as well as the $C$ (Wilson line) moduli. Various limits of strong 
coupling -- large radii are examined in detail. Section 6, finally 
summarizes our results and gives an outlook. The appendix contains some
useful technical details.

\sect{The relation between $M_{\rm Planck}$ and
$M_{\rm string}$ at one--loop}

As a starting point we write the N=1 effective action for 
the heterotic string
through a superconformal Lagrangian as $D$--density
with a gauge invariant linear multiplet\footnote{For more details
see \cite{dqq}. This reference contains  a  complete list
of references.} $\hat L$ (without superpotential)
\cite{cecotti,dfkz,agnt1}:

\be\label{super}
\Lc_{linear}=\lf\{S_0\ov S_0\Phi\lf(\fc{\hat L}{S_0\ov S_0},\Si,\ov \Si\ri)
\ri\}_D
\ee
and

\be
\Phi\lf(\fc{\hat L}{S_0\ov S_0},\Si,\ov \Si\ri)=
-\fc{1}{\sqrt 2} \lf(\fc{\hat L}{S_0\ov S_0}\ri)^{-\h} e^{-G^{(0)}/2}
-\h \fc{\hat L}{S_0\ov S_0}\fc{G^{(1)}}{16\pi^2}\ ,
\ee
where the chiral superfield $S_0$ denotes the compensator and the chiral
superfields $\Si=(z,\psi_z,f_z)$ refer to the moduli fields 
$z=T,U,C,\ldots $ describing the vacuum of the underlying string model.
The moduli dependent functions $G^{(0)}$ and 
$G^{(1)}$ will be specified later.
Note, that all non--holomorphic gauge 
dependence is encoded in the $D$--density \cite{dfkz}.
When fixing dilatation symmetry, we may arrive at different actions
depending on how we choose the vev of the scalar component $z_0$ 
of the compensator. For the Einstein frame we take \cite{dqq}

\be
(z_0\ov z_0)^{3/2}=\sqrt 2 \kappa^{-2} c^\h e^{G^{(0)}/2}
\label{fixing}
\ee
and the resulting bosonic Lagrangian at one--loop becomes:


\bea\label{actionlinear}
\ds{e^{-1}\Lc_{\rm Einstein}}&=&\ds{-\fc{1}{2\kappa^2}R-\fc{1}{4\kappa^2}
\fc{1}{c^2}\p_\mu c\p^\mu c+\fc{1}{4\kappa^2}\fc{1}{c^2} v_\mu v^\mu}\nnn
&-&\ds{\kappa^{-2}\lf[G^{(0)}_{z\ov z}-\fc{\kappa^2c}{16\pi^2} 
G^{(1)}_{z\ov z}\ri]\p_\mu z\p^\mu\ov z-\fc{1}{4}
\lf[\fc{1}{\kappa^2 c}-\fc{G^{(1)}}{16\pi^2}\ri]F^a_{\mu\nu}F^{a\mu\nu}\ .}
\eea
Here 
$c$ is the lowest component of the linear multiplet. The vector 
field $v_\mu$ obeys the constraint $\p_\mu v^\mu=0$.
There are no one--loop corrections to the Einstein term
\cite{earlyanton,kk}, to the kinetic terms for field $c$ 
and the antisymmetric tensor $b_{\mu\nu}$ which is
contained in $v_\mu$ via $v_\mu\sim\ep_{\mu\nu\rho\si}\p^\nu b^{\rho\si}$
\cite{ms2}.
Notice that $\kappa^2 c$ is the
string--loop counting parameter, $G^{(0)}$ is the tree--level 
part of the K\"ahler potential for the fields $z$ 
and $G^{(1)}$ denotes the one--loop corrections to it 
\cite{dfkz,agnt1,agnt2,afgnt,pet}.

Before we extract from this Lagrangian model independent relations,
let us perform a duality transformation on \req{super} to 
eliminate the linear multiplet $\hat L$ by 
introducing an additional chiral multiplet $S$. Then 
\req{super} involves only chiral fields:

\be\label{actionchiral}
\Lc_{chiral}=-\fc{3}{2}
\lf\{S_0\ov S_0\lf[e^{-G^{(0)}/3}\lf(\fc{-iS+i\ov S}{4}+
\h \fc{G^{(1)}}{16\pi^2}\ri)^{1/3}\ri]\ri\}_D
+\fc{1}{4}\{\fc{-iS}{2} W^a W^a\}_F\ .
\ee
The scalar kinetic terms of the action \req{actionchiral} 
with the fixing 

\be\label{FIX}
z_0\ov z_0=\kappa^{-2} \lf(\fc{-iS+i\ov S}{4}+
\h \fc{G^{(1)}}{16\pi^2}\ri)^{-1/3} e^{G^{(0)}/3}\ ,
\ee
corresponding to \req{fixing} follow then from the K\"ahler potential:

\be\label{kaehler}
K=-\kappa^{-2}\ln\lf[-iS+i\ov S+\fc{1}{8\pi^2}G^{(1)}(z,\bar z)\ri]
-\kappa^{-2} G^{(0)}(z,\bar z)\ .
\ee
The bosonic terms are

\bea\label{lowaction}
\ds{e^{-1}\Lc_{\rm Einstein}}&=&
\ds{-\fc{1}{2\kappa^2}R+\fc{\kappa^{-2}}{(-iS+i\ov S+\fc{1}{8\pi^2}G^{(1)})^2}
\p_\mu S\p^\mu\ov S-\fc{1}{4}\lf[\fc{-iS+i\ov S}{2}\ri]
F^a_{\mu\nu}F^{a\mu\nu}}\nnn
&-&\ds{\kappa^{-2}\lf[G^{(0)}_{z\ov z}+
\fc{G^{(1)}_{z\ov z}/8\pi^2}{-iS+i\ov S+\fc{1}{8\pi^2}G^{(1)}}-
\fc{G^{(1)}_zG^{(1)}_{\bar z}/(8\pi^2)^2}{(-iS+i\ov S+\fc{1}{8\pi^2}
G^{(1)})^2}\ri]\p_\mu z\p^\mu\ov z\ ,}\nnn
&-&\ds{\kappa^{-2}\lf[\fc{iG^{(1)}_{\bar z}/8\pi^2}{(-iS+i\ov S+
\fc{1}{8\pi^2}G^{(1)})^2}\p_\mu S\p^\mu\ov z-
\fc{iG^{(1)}_{z}/8\pi^2}{(-iS+i\ov S+
\fc{1}{8\pi^2}G^{(1)})^2}\p_\mu z\p^\mu\ov S\ri]\ .}
\eea
As we see from \req{kaehler} the function 
$G^{(1)}$ amounts to an --in general-- 
non--holomorphic shift of the dilaton field $S$.
We shall return to this shift (and possible further
holomorphic shifts originating from the $F$--density) in section 3.1.
Note that the linear formalism \req{actionlinear} is the most natural one
to discuss non--harmonic gauge
couplings given by $G^{(1)}$ \cite{dfkz}.

Now we can compare 
eq. \req{actionlinear} with \req{lowaction}.
Let us first extract the tree--level relation

\be\label{diltree}
\lf. g_{\rm string}^{-2}\ri|_{bare}=\fc{1}{\kappa^2 c}
\stackrel{duality}{=}\fc{-iS+i\ov S}{2}\ ,
\ee
by looking at the kinetic term for $c$ or $S$ and setting $G^{(1)}=0$.
Inspection of the gauge terms in \req{lowaction}
leads to the convention:

\be\label{S}
S=\fc{\th_a}{8\pi^2}+i \fc{1}{g_a^{2}}\ .
\ee
With the identification $M_{\rm string}^2=\lng c\rng$ and 
$\kappa^{-1}=M_{\rm Planck}$ we may cast \req{diltree} into

\be\label{pss0}
M_{\rm Planck}^2=\im(S)\ M_{\rm string}^2\ ,
\ee
or: 

\be\label{pss}
M^2_{\rm Planck}=\lf. g^{-2}_{\rm string}\ri|_{bare}\ M^2_{\rm string}\ .
\ee
This is the well--known relation for the heterotic string \cite{ghmr}
at tree--level.
At one--loop one obtains the following string--coupling:

\be\label{oneres}
\lf.g_{\rm string}^{-2}\ri|_{one-loop}=\fc{1}{\kappa^2 c}
\stackrel{duality}{=}\fc{-iS+i\ov S}{2}+\fc{1}{16\pi^2}G^{(1)}\ .
\ee
Again, the last equality follows from the duality transformation. 
Here the field
$S$ appears as a Lagrange multiplier and has to be defined
order by order in perturbation theory.
The linear multiplet has a fixed relation to string vertices
and therefore the left hand side of \req{oneres} stays invariant under 
all perturbative symmetries.

We want to elaborate whether the relation \req{pss}
is stable against one--loop corrections.
As a consequence of eq. \req{actionlinear} the form\footnote{We rescale
fields so that we arrive at canonical kinetic energy terms, e.g.: 
$g_{\mu\nu}=2\kappa h_{\mu\nu}$.} of the coupling
of the graviton to the dilaton energy momentum tensor 
(with $\kappa^2 c=e^{2D}$)

\be
\kappa h_{\mu\nu}(\p^\mu D\p^\nu D-\eta^{\mu\nu}\p_\al D\p^\al D)
\label{coup}
\ee
remains unchanged at the one--loop level,
since the kinetic terms of the field $c$ and the graviton do not change.
To clarify whether \req{pss} receives one--loop corrections 
we have to extract from the string one--loop 
amplitude\footnote{In \cite{ghmr}, a comparison of \req{coup}
and the tree--level string amplitude \req{kinetic} lead to the 
conclusion \req{pss}.} $\Ac[D(k_1),G(h_{\mu\nu}),D(k_3)]$
a possible contribution to the term 

\be\label{kinetic}
\lf. g_{\rm string}\ri|_{bare} 
\ep^2_{\mu\nu}(k^\mu_1k^\nu_3+k^\nu_1k^\mu_3)\ .
\ee
Here $\ep_{\mu\nu}^2$ is the polarization tensor of the graviton.
The coupling $\lf. g_{\rm string}\ri|_{bare}$
appears as normalization of the graviton vertex
operator, which is given by the non--linear sigma--model action.
Therefore it should be taken at tree--level. Since it is the
linear multiplet, which has a fixed relation to the dilaton vertex operator,
we are really determing the one--loop correction to \req{coup} rather than 
to $\kappa h_{\mu\nu}(\p^\mu S\p^\nu S-\eta^{\mu\nu}\p_\al S\p^\al S)$.
This term can be worked out easily by using \cite{fotw}. 
There it was shown
that the $O(k^2)$ part of a three graviton amplitude vanishes.
We have only to replace two graviton polarization tensors with those 
of two dilatons and work out the contribution corresponding to \req{kinetic}
to see that it vanishes for the same reasons.
However, this shows that the relation \req{pss}
is unchanged at one--loop.

To summarize,  from \req{actionlinear} one deduces that the 
form for the coupling of the graviton to the dilaton energy momentum 
tensor does not change at one--loop 
and by doing an explicit string one--loop calculation one verifies  
that \req{pss} is {\em not} changed\footnote{
Related discussions may be found in \cite{kkkk}.} at one--loop.
Nevertheless, the meaning of $S$ does change at one--loop
when going from \req{diltree} to \req{oneres} and
the right hand side of \req{pss0} refers to quantities defined at
tree--level. 
Therefore, at one--loop, the right hand side of eqs. 
\req{pss0}, \req{pss} should better be written entirly
in terms of one--loop quantities:

\be\label{finalgross}
M_{\rm Planck}^2=\lf[\im(S)+\fc{1}{16\pi^2}G^{(1)}\ri]\ M_{\rm string}^2\ .
\ee

\sect{String--unification and low--energy predictions}
\subsect{String threshold corrections}

At string tree level the gauge couplings, encoded 
in \req{S}, are related to the string coupling \req{diltree}
via \cite{gin87} 

\be\label{hyper}
g^{-2}_a=k_a \lf. g_{\rm string}^{-2}\ri|_{bare}=k_a\im(S)\ .
\ee
Here $k_a$ is the Kac--Moody level of the group factor labeled by $a$.
At one--loop, the gauge coupling in eq. \req{actionlinear}, which
follows from the $D$--density \req{super}
receives additional harmonic 
contributions $\triangle_{\rm harmonic}$ from integrating out heavy 
string states. These corrections are usually summarized in the $F$--density.
Besides non--harmonic pieces $\triangle_{\rm triangle}$
arising from triangle graphs involving massless fields and some 
constants $c_a$. 
The effective gauge coupling at the scale $\mu=M_{\rm string}$ then reads:

\bea
\ds{g_{a,eff.}^{-2}}&=&\ds{\fc{1}{\kappa^2c}-\fc{k_a}{16\pi^2}G^{(1)}+
\fc{1}{16\pi^2}(\triangle_{\rm triangle}+\triangle_{\rm harmonic})+c_a}\nnn
&=&\ds{\lf. g_{\rm string}^{-2}\ri|_{one-loop}
+\fc{1}{16\pi^2}\triangle_a+c_a\ .}
\label{triangle}
\eea
In \req{actionlinear} $g_{a,eff.}^{-2}$ then appears as the bare coupling 
in front of $F^2$.
The one--loop corrections to the physical gauge coupling are denoted 
by $\triangle_a$.

In the following we want to discuss toroidal orbifold models 
with $d=4$, N=1 space--time supersymmetry \cite{dhvw}.
They provide an interesting class of realistic string models 
and their moduli--space is rich enough to discuss various
aspects of string theory. Their spectrum may contain a subsector
which can be arranged in N=2 multiplets. For this subsector
all perturbative corrections like $\triangle_a,G^{(1)}$ 
can be calculated.
In these models $\triangle_a$ can be split into three pieces:

\be\label{ddd}
\triangle_a=\al_a\triangle-k_aG^{(1)}+k_a\si\ .
\ee 
This splitting makes sense, since we know from \req{actionlinear}, that
the one--loop correction $\triangle_a$ contains a non--holomorphic
piece $G^{(1)}$ which has to cancel the mixing between the dilaton
and the moduli fields given in \req{oneres}.
The latter contains so--called GS--corrections\footnote{These terms
cancel target--space anomalies  arising from 
triangle graphs with a coupling to the K\"ahler-- and sigma--model connections
in four dimensions.
These corrections
should not be confused with the anomaly cancelation terms in ten dimensions,
which we will discuss in sect. 5. Their job
is to cancel gauge and gravitational anomalies.}
$G^{(1)}_{N=1}$ 
referring to twisted planes and therefore involving
only moduli of these planes \cite{dfkz}.
Furthermore, one has GS--corrections $G^{(1)}_{N=2}$ coming from the 
untwisted planes.
These corrections give rise to IR--divergent wave function 
renormalizations due to singularities associated with additional massless
particles at subvarieties of the moduli space.
The generic gauge--group dependent moduli--dependent threshold--corrections
are contained in $\triangle$. They have been first calculated in
\cite{DKL2} for orbifold models without Wilson lines
and in \cite{ms5,k1} for orbifold models with one Wilson line.
Generalization of the latter to more Wilson lines have been performed
in \cite{hm,mm}.  
The coefficient $\al_a$ 
is the total anomaly coefficient for the
sigma--model and K\"ahler connection anomaly.
Only matter which fits in N=2 multiplets
contributes to $\triangle$. The following splitting seems to be
convenient \cite{kl2}

\be\label{splitting}
\fc{b_a^{j,N=2}}{|D|/|D_j|}=\al^j_a-k_a\delta^j_{GS}\ ,
\ee
(with $b_a^{j,N=2}$ being the N=2 $\beta$--function coefficient), 
since the anomaly $\al^j_a$ 
with respect to the orbifold plane $j$ is 
partially cancelled by the string thresholds $b_a^{j,N=2}\Delta$ and a 
GS--term $k_a\delta^j_{GS}\Delta$, respectively \cite{dfkz}.
Here $D,D_j$ refer to the orbifold group and to a subgroup
generating the N=2 sector, respectively.

To illustrate this mechanism, let us give an example.
For orbifolds with untwisted planes $j$ the equation \req{splitting}
turns out to be fulfilled in such a way that $\delta_{GS}^j=0$  \cite{kl2}.
Therefore, we choose the standard $Z_3$--orbifold with gauge group 
$SU(3)\times E_6\times E_8'$. In this case one has no fixed orbifold plane. 
Thus, we have $b_a^{j,N=2}=0$ and we need a GS--term 

\be
G^{(1)}_{N=1}=\sum_{j=1,2,3}\delta^j_{GS} \ln(-iT^j+i\ov T^j)\ ,
\ee
to cancel the total anomaly, given by 
$\al^j_a\ln(-iT^j+i\ov T^j)$ with $\al^j_a=-30$ for the plane $j$
and $\delta^j_{GS}=-30$.
Although in \req{triangle} the whole
moduli--dependence disappears for these sectors, i.e. $\triangle_a=0$,
in the combination of \req{ddd} the correction 

\be
\si_{N=1}(T^k,U^k)=-\sum_{j=1,2,3}\de^j_{GS} \ln |\eta(T^j)|^4 |\eta(U^j)|^4\ .
\ee
may be thought as an universal correction.
To keep the string--coupling \req{oneres} invariant,
the dilaton field $S$ has to transform properly. We may use 
the universal correction $\si$ to perform a holomorphic field redefinition
of the dilaton field $S$

\be
-iS_{\rm inv.}:=-iS-\fc{1}{8\pi^2}
\sum_{j=1,2,3}\delta^j_{GS} \ln\eta^2(T)\eta^2(U)
\ee
to obtain an invariant field $S_{\rm inv.}$ \cite{dfkz}.
This is the only possible source for moduli dependent 
universal corrections $\si_{N=1}$ appearing from
twisted planes.
 
Let us now turn to the GS--type corrections $G^{(1)}_{N=2}$ arising from an
untwisted plane $T^2$ with moduli $T,U$ and possible Wilson line modulus
$C$. These corrections can be expressed by the N=2 prepotential:
thanks to special geometry it can be entirely written as derivatives of the
prepotential \cite{dewit,afgnt,hm}

\be\label{mixing}
G^{(1)}_{N=2}=-\fc{32\pi^2}{Y}\re\lf[h^{(1)}-\h(T-\bar T)\p_T h^{(1)}
-\h(U-\bar U)\p_U h^{(1)}-\h(C-\bar C)\p_C h^{(1)}\ri]\ ,
\ee
with [cf. the K\"ahler potential in \req{kaehler}]:
\be\label{mixingY}
Y=e^{G^{(0)}}=-(T-\ov T)(U-\ov U)+\fc{1}{4}(C-\ov C)^2\ .
\ee
The group--independent corrections $\si_{N=2}$, which do not contain 
the one--loop K\"ahler correction $G^{(1)}$, are related to
the charge insertion which appears in the threshold calculation and
the gravitational back reaction to the 
background gauge fields in \cite{vk}. They have been first derived in 
\cite{kk}, further developed in \cite{petriz,kkkk} and 
will be generalized to more moduli in section 4.
In general, there are also moduli--independent 
corrections $c_a$ from N=1 sectors and from the different IR--regularization 
of field-- and string--theory. But these are small
and will be neglected in the following. They are discussed in 
\cite{vk,fermionic}.

\subsect{Perturbative string--unification}

We go to a string model with N=1, $d=4$ space--time supersymmetry
and the gauge group of the Standard Model $SU(3)\times SU(2)\times U(1)$.
Such models can be constructed as toroidal orbifolds with Wilson lines.
The Wilson lines break the gauge groups and may also reduce their rank. 
For more details see refs. \cite{hpn1,wend}.
The corrections in \req{triangle}
spoil the string tree--level result \req{hyper} and 
split the one--loop gauge couplings at
$M_{\rm string}$.
This splitting could allow for an effective unification at a scale 
$M_{\rm GUT} <M_{\rm string}$ or destroy the unification.
The identities \req{triangle} serve as boundary conditions for
the running field--theoretical couplings valid below \ms.
The evolution equations below $M_{\rm string}$ 

\be\label{evo}
\fc{1}{g_a^2(\mu)}=k_a\im(S)+\fc{b_a}{16\pi^2} 
\ln \fc{M_{\rm string}^2}{\mu^2}+\fc{1}{16\pi^2}\al_a\triangle
+\fc{k_a}{16\pi^2}\si\ ,
\label{running}
\ee
allow us to determine $\sin^2 \th_{\rm W}$ and
$\al_{\rm S}$ at $M_Z$. 
After eliminating $\im(S)$ in the second and third equation 
one obtains

\bea\label{mz}
\sin^2\th_{\rm W}(M_Z)&=&\ds{\fc{k_2}{k_1+k_2}-
\fc{k_1}{k_1+k_2}\fc{\al_{em}(M_Z)}{4\pi}
\lf[\Ac\ln\lf(\fc{M_{\rm string}^2}{M_Z^2}\ri)+\Ac'\triangle\ri]\ ,}\nnn
\al_{S}^{-1}(M_Z)&=&\ds{\fc{k_3}{k_1+k_2}\lf[\al_{em}^{-1}(M_Z)
-\fc{1}{4\pi}\Bc
\ln\lf(\fc{M_{\rm string}^2}{M_Z^2}\ri)-\fc{1}{4\pi}\Bc'\triangle\ri]\ ,}
\eea
with  $\Ac=\fc{k_2}{k_1}b_1-b_2, \Bc=b_1+b_2-\fc{k_1+k_2}{k_3}b_3$ and 
$\Ac'=\fc{k_2}{k_1}\al_1-\al_2$
and $\Bc'=\al_1+\al_2-\fc{k_1+k_2}{k_3}\al_3$. 
For the MSSM one has $\Ac=\fc{28}{5},
\Bc=20$. The coefficients $\Ac',\Bc'$ depend on the string model under
consideration, i.e. how its relevant particle content enters
the anomaly coefficients in \req{splitting} \cite{IL}.
These two equations determine
 the gauge group dependent threshold corrections 
$\triangle$ and $M_{\rm string}$ that are necessary
to obtain the correct experimental low--energy data $\sin^2_{\rm W}(M_Z)$ and 
$\al_{\rm S}(M_Z)$.
The three couplings meet at the single point

\be\label{meet}
M_{\rm GUT}=M_{\rm string}\ e^{\fc{\Ac'}{2\Ac}\triangle}\sim 
2 \cdot 10^{16}\gev\ ,
\ee
if the following relation holds:

\be
\Ac'\Bc=\Ac \Bc'\ .
\ee
Note, that the universal correction $\si$ does not play any r\^ole when
considering the low--energy implications \req{mz} and \req{meet}.
However it does influence the unification coupling
constant e.g.: $k_2^{-1}g_2^{-2}(M_{\rm GUT})$:

\be\label{gutstring}
k_2^{-1}g_2^{-2}(M_{\rm GUT})=\im(S)+\fc{1}{16\pi^2}\si
+\fc{1}{16\pi^2k_2}
\lf(\al_2-b_2\fc{\Ac'}{\Ac}\ri)\Delta\ .
\ee
The last term vanishes\footnote{In the model $(\al_2=5/2)$ 
considered in \cite{ns}: 
$\Ac'\triangle\sim 2(-16)$, i.e. $-\fc{1}{16\pi^2}\fc{b_2}{k_2}
\fc{\Ac'}{\Ac}\Delta+\fc{1}{16\pi^2}\al_2\Delta\sim -0.2$ leading to
$g_2^{-2}(M_{\rm GUT})=1.9$.}  
 for $\al_a=b_a$.
Therefore we define:

\be\label{defi}
g^{-2}_{\rm GUT}:=\im(S)+\fc{1}{16\pi^2}\si\ .
\ee
Thus, we expect $\im(S)+\fc{1}{16\pi^2}\si\sim 2.0$.
There is yet another reason to see this: Combine the first and second eqs.
of \req{running} to:

\be\label{alem}
g_{\rm GUT}^{-2}\equiv
\im(S)+\fc{1}{16\pi^2}\si=\fc{1}{k_1+k_2}\lf[\fc{1}{4\pi}
\al_{em}^{-1}(M_Z)-\fc{1}{16\pi^2}(b_1+b_2)\ln \fc{M_{\rm string}^2}{M_Z^2}
-\fc{\al_1+\al_2}{16\pi^2}\Delta\ri]\ .
\ee
For the model $(\al_1+\al_2=10)$ considered in  \cite{ns} one 
immediately obtains for the right hand
side the value $2.1$. On the other hand, looking at the solutions
of \req{alem} one realizes: 

\be
1.9\leq \im(S)+\fc{1}{16\pi^2}\si\leq 2.1\ .
\ee
arising from the two cases $\triangle =0;\ M_{\rm string}=2\cdot 10^{16}\gev$
and $\triangle=-16;\ M_{\rm string}=4\cdot 10^{17}\gev$
which both solve eqs. \req{mz}. Therefore we conclude: 

\be\label{gutt}
g^{-2}_{\rm GUT}\sim\im(S)+\fc{1}{16\pi^2}\si\sim 2\ .
\ee

After a rearrangement of some scheme dependent parts in \req{running}
and one obtains\footnote{The  
relation between the quantities $\si, G^{(1)}$ and $Y$ defined in 
\cite{kk,petriz,kkkk} is: $-Y=-G_{N=2}^{(1)}+\si$. In the next sect.
we will see that the correction $G^{(1)}$ is rather small compared to
$\si$. Actually, $G_{N=2}^{(1)}\sim 1/R^2$ with $R$ being the 
compactification radius. Therefore in the following we will concentrate our
discussion on the correction $\si$.} from \req{finalgross}

\be\label{schluss}
M_{\rm string}= 0.527 \cdot g_{\rm GUT} \cdot 10^{18} \gev  
\fc{1}{\sqrt {1-\fc{g_{\rm GUT}^2}{16\pi^2} [\si-G^{(1)}]}}\ .
\ee

Therefore the determination of the correct  solution 
$(\Delta,M_{\rm string})$ of \req{mz}
requires the knowledge of either $\si$ or 
the vev of the dilaton $S$. 
Let us present some solutions to \req{mz} for the model
discussed in \cite{ns}.

\bdm
\begin{tabular}{|c||c|c|c|c|c|c|}
\hline
$\triangle$     &   0  &$-5$& $-10$ &   $-16.75$ & $-20$&$-30$ \\[2mm]
\hline
$M_{\rm string} [10^{17} Gev]$&  0.2 &0.44& 1.1&   3.6      & 6.4&38.5  \\[2mm]
\hline
$\si$   &$-133121$&$-22108$ & $-3474$      &  0&237&347      \\[2mm]
\hline
\end{tabular}
\edm

\bc
{\small \em Tab.1 -- Solutions $(\triangle,M_{\rm string})$
of eq. \req{mz} which require after \req{schluss} a specific $\si$.}
\ec
\ \\
From eq. \req{schluss} one learns that only  huge 
values $\si\sim O(-10^5)$ would influence this equation, 
i.e. may considerably lower 
the string--scale down to the GUT--scale \req{meet}.
It is the smallness of $g_{\rm GUT}^2/16\pi^2\sim 10^{-3}$ 
entering the formula \req{schluss} that is responsible for
this fact.
Since both $\si$ and $\Delta$ are moduli--dependent functions
it is the vev of the moduli fields, which finally selects
one of the above solutions.  
In \cite{ns} we investigated the moduli dependences of $\triangle$
and its influence to string unification. We assumed the term 
$\si/16\pi^2$ to be small and found 
that the solution $(\triangle=-16.75,\sigma=0)$ can be achieved
with rather small vevs of the moduli fields: $T/2\al'=4i=U$ and $C=\sqrt 2/4$.
In the next section we will elaborate the full moduli dependence
of $\si(T,U,C)$. 
We shall see that
in the perturbative regime
the correction $\si$  is always negative, i.e. the effect of $\si$
 results in a lowering of the string--scale. 
As a consequence of \req{gutt} this  drives the string coupling
to smaller values. 
A huge negative $\si$ requires a huge vev of $T$ which on the other hand 
pushes $\triangle$ into the opposite direction. In the next section,
we will see, that with $\sigma(T,U,C)$ we cannot reach such large values 
as required from Tab.1. 
At $T=4i=U,\ C=\sqrt 2/4$, for example,
 we have $\Delta\sim -17$ and $\si\sim -50$. Although both are
of comparable size\footnote{With $\Ac'\sim 2$, i.e. 
$\Ac'\Delta\sim -36$ these two values enter \req{evo} as corrections of
the same order.}, only $\Delta$ is relevant for unification.
In general, the preferred solution to \req{mz} is  
$\triangle=-16.75$, achieved with $T/2\al'=4i=U$ and $C=\sqrt 2/4$.
The universal corrections $\si$ do not play 
an important r\^ole in string unification.
Nevertheless, in section 5 we will see that they become important
when moving to stronger coupling.

\sect{Universal one--loop corrections}

In this section we want to derive the Wilson line dependence 
of the universal one--loop correction $\si$. Its moduli dependence
is completely given by the N=2 subsectors of the full 
N=1 $d=4$ heterotic string theories we have discussed in the previous section.
For concreteness we will do this calculation for an $N=1$ $d=4$ 
toroidal orbifold
which has an N=2 subsector, generated by a $Z_2$ orbifold twist
which leaves one torus $T^2$ fixed and leads to the N=2 gauge group
$SU(2)\times E_7\times E_8'$. In addition we introduce a Wilson line 
($C$) with respect to
to the torus $T^2$, which may break  $E_7$ down to $SO(12)$ 
for nontrivial vevs of $C$. The torus is described by the two moduli $T$
and $U$.
This example should then allow us to study
all the relevant properties of the N=1 models introduced above,
including the Higgs mechanism
for the N=1 gauge group. For more details see \cite{wend}.

\subsection{Without Wilson line}

To warm up let us first consider the case without a Wilson line.
This case has already been discussed in quite detail in \cite{kk,petriz,kkkk}.
Nonetheless we would like to repeat the calculation here because of
the following  two reasons.
By using \cite{hm} we find an alternative way to obtain these results
by means of a differential equation for the prepotential.
Secondly, this method seems to be more convenient for cases involving
an arbitrary number of Wilson lines of which we will make use  it in 
section 4.2.
In this model we may consider the two physical gauge groups 
$G_a=E_7,\ E_8'$. 
For them we do not expect any singularities in the $(T,U)$--moduli space
and the gauge group dependent part $\triangle$ of their threshold corrections
$\triangle_a$ \req{ddd}
are expressed by the well--known formula \cite{DKL2}:

\be\label{logT}
\triangle=-\ln\lf[(-iT+i\ov T)(-iU+i\ov U)|\eta(T)|^4|\eta(U)|^4\ri]\ .
\ee
The full correction $\triangle_a$ enters in a second order differential 
equation for the one--loop correction  $h^{(1)}$ to the prepotential 
of the underlying N=2 theory \cite{hm}

\be\label{dgl0}
\re\lf[8\pi^2\p_T\p_Uh^{(1)}(T,U)\ri]-
G_{N=2}^{(1)}=\triangle_a(T,U)+4\re\lf[\ln\Psi_a(T,U)\ri]+
b^{N=2}_aK_0\ ,
\ee
with:

\be
\Psi_a(T,U)=[\eta(T)\eta(U)]^{b_a^{N=2}}[j(T)-j(U)]^{1/2}\ \ ,\ \ 
b_{E_7}^{N=2}=84,\ b^{N=2}_{E_8'}=-60\ .
\ee
Using \req{dgl0} and \req{ddd} we are able to extract $\si(T,U)$ \cite{dewit}

\be\label{cindy0}
\si(T,U)=\re\lf\{8\pi^2\p_T\p_Uh^{(1)}(T,U)-2\ln[j(T)-j(U)]\ri\}\ ,
\ee
which gives for the prepotential in the chamber $T_2>U_2$ \cite{hm}

\be\label{prep0}
h^{(1)}(T,U)=-\fc{i}{12\pi}U^3-\fc{1}{(2\pi)^4}
\sum_{(k,l)>0}c_1(kl)Li_3[e^{2\pi i (kT+lU)}]+const.
\ee
with $Li_3(x)=\sum_{k=1}^\infty x^k/k^3$ and $\fc{E_4E_6}{\eta^{24}}=
\sum_{k=-1}^\infty c_1(k) q^k$  finally resulting in

\be\label{final0}
\si(T,U)=\re\lf\{-2\ln[j(T)-j(U)]-2\sum_{(k,l)>0}c_1(kl)\ kl
\ln\lf[1-e^{2\pi i(kT+lU)}\ri]\ri\}\ .
\ee
In fact, the large $T$--limit leads to

\be
\si(T,U)\lra -4\pi\im(T)\ \ \ ,\ \ \ T\lra i\infty\ ,
\ee
in agreement\footnote{One has to multiply by a factor of $-3/2$
to take into account the three orbifold planes and the N=1
structure of their $Z_2\times Z_2$ orbifold. The minus sign arises from the
different conventions for $\si$ in \req{ddd}.}
with the limit of \cite{petriz}.
As expected for the threshold--corrections
under consideration, $\si(T,U)$ stays finite throughout the moduli space.
In particular in the limit $T\ra U$ the contribution of $k=1,\ l=-1$
in the sum cancels precisely the first term in \req{final0}.
The combination $\im(S)+\fc{1}{16\pi^2}\si(T,U)$ 
is invariant under the perturbative 
duality group $SL(2,Z)_T\times SL(2,Z)_U\times Z_2^{T\leftrightarrow U}$.
However, $\si$ alone, receives shifts from $Z_2^{T\leftrightarrow U}$.
On the other hand, that behaviour precisely compensates the non--invariance
of $S$ arising from its multi--valuedness.

\subsection{Wilson--line dependence}

Now we go to the case with 
non--trivial gauge background fields of an $E_7$ 
subgroup.
More concretely, we turn on a Wilson line in an $SU(2)$ subgroup
of $E_7$. For a generic vev of the Wilson line modulus $C$
the gauge group $E_7$ is broken to $SO(12)$.
This then leads to the phenomenological interesting case of
 physical gauge couplings which develop logarithmic singularities
in certain regions of the moduli space $(T,U,C)$, when particles 
charged with respect to  the gauge group under consideration, i.e. $SO(12)$,
become massless.
The gauge group--dependent part $b_a^{N=2}\Delta$ of one--loop
corrections with a Wilson line modulus were first derived in \cite{ms5} 
by looking at their perturbative modular symmetries and their 
singularity structure in the moduli space. 
Two cases of physical gauge couplings are discussed.
In the first case, referring to $E_8'$,
no particles become massless for $C\ra 0$ and
the form of these thresholds is given by\footnote{In appendix A some
basics of Siegel modular forms are presented. The relevance of
Siegel modular forms in the context of string one--loop corrections
was originally observed in \cite{ms5}.} \cite{ms5}:

\be\label{d2}
\Delta^{II}=-\fc{1}{12}\ln(\kappa Y^{12})|\chi_{12}(\Om)|^2\ .
\ee
In the second case some particles,
charged under both the $SU(2)$ and $SO(12)$, become massless
for $C\ra 0$. This means that the effective one--loop correction 
develops a logarithmic singularity in this limit since these particles
run in the loop. The form of these thresholds is given by:

\be\label{d1}
\Delta^I=-\fc{2}{5}
\ln\lf|\fc{\chi_{10}^{1/2}(\Om)}{\chi_{12}^{5/12}(\Om)}\ri|^2-
\fc{1}{12}\ln(\kappa Y^{12})|\chi_{12}(\Om)|^2\ .
\ee
The quantity $Y$ has been defined in \req{mixingY}.
The number $2/5$ refers to the additional contribution (relative to
$b_{SO(12)}^{N=2}$) to the $\beta$--function arising from the particles
becoming massless.
No universal contributions are included in these functions,
as they refer to $\triangle$ in \req{ddd}.
This means that they can be determined by considering the difference of two 
gauge groups \cite{stieberg}.

Also here we can now derive a differential equation
for the one loop correction $h^{(1)}$ to the prepotential $h$

\be\label{prep}
h(S,T,U,C)=-iS(TU-C^2)+h^{(1)}(T,U,C)+O(e^{2\pi i S})\ ,
\ee
with \cite{k2}

\be
h^{(1)}(T,U,C)=-\fc{i}{4\pi}d(T,U,C)-\fc{1}{(2\pi)^4}
\sum_{(k,l,b)>0}c_1\lf(kl-\fc{1}{4}b^2\ri)Li_3[e^{2\pi i (kT+lU+bC)}]+const.\ ,
\ee
$\fc{E_4\hat{E}_{6,1}}{\eta^{24}}=
\sum\limits_{k\in \Z,\ k\in \Z+3/4}c_1(k)q^k$ and
$(k,l,b)>0$ is composed by the three 
orbits $(i)\ k>0,l\in \Z, b\in \Z,$\ 
$(ii)\ k=0, l>0, b \in \Z,$\ 
$(iii)\ k=0=l, b<0$.  
Moreover, 

\be\label{cubic}
d(T,U,C)=\fc{1}{3} U^3+\fc{40}{3}C^3-7 UC^2-6TC^2\ .
\ee
There are ambiguities for the cubic polynomial \req{cubic} due to the
fact that the holomorphic prepotential is only fixed up to quadratic pieces
in the homogeneous coordinates $\hat X^I$.
These quadratic pieces include e.g. cubics in $C$. On the other hand,
this ambiguity can be fixed when comparing the prepotenial
with the corresponding one of the typeII theory, which leads to
the form \req{cubic} \cite{peter,k2,ccl}.
For the differential equation one has \cite{stieberg}:

\be\label{dgl}
\re\lf[\fc{32\pi^2}{5}(\p_T\p_U-\fc{1}{4}\p^2_C)h^{(1)}\ri]-
G_{N=2}^{(1)}=\triangle_a+\fc{16}{5}\re\lf[\ln\Psi_a(T,U,C)\ri]+
b^{N=2}_aK_0\ .
\ee
This differential equation holds for both types of gauge groups 
$G_a=E_8',\ SO(12)$ with 

\bea\label{beta}
\ds{b_{SO(12)}^{N=2}}&=&\ds{60,}\nnn
\ds{b_{E_8'}^{N=2}}&=&\ds{-60}
\eea
and the functions

\bea\label{psis}
\ds{\Psi_{SO(12)}(T,U,C)}&=&\ds{[4i\chi_{35}(\Om)]^{1/2}
[-4\chi_{10}(\Om)]^{2}\ ,}\nnn
\ds{\Psi_{E_8'}(T,U,C)}&=&\ds{\fc{[4i\chi_{35}(\Om)]^{1/2}}
{[-4\chi_{10}(\Om)]^{11/2}}\ ,}
\eea
respectively.
Using \req{dgl} and \req{ddd} we can extract $\si(T,U,C)$

\be\label{cindy}
\si(T,U,C)=\re\lf\{\fc{32\pi^2}{5}(\p_T\p_U-\fc{1}{4}\p^2_C)h^{(1)}(T,U,C)
-\fc{4}{5}\ln\lf(\fc{\Delta_{35}^2(\Om)}{\Delta_{10}^{22}(\Om)}\ 
\chi_{12}^{25/2}(\Om)\ri)\ri\}\ ,
\ee
with $\Delta_{35}=4i\chi_{35}$ and $\Delta_{10}=-4\chi_{10}$ 
which gives for the chamber $T_2>U_2>2 C_2$:

\bea\label{final}
\si(T,U,C)&=&\ds{\re\lf\{-\fc{8\pi i}{5}(3T+\fc{7}{2}U-20C)
-\fc{4}{5}\ln\lf(\fc{\Delta_{35}^2(\Om)}{\Delta_{10}^{22}(\Om)}\ 
\chi_{12}^{25/2}(\Om)\ri)\ri.}\nnn
&-&\ds{\lf.\fc{4}{5}\sum_{\al=(k,l,b)>0}\al^2 c_1\lf(\fc{\al^2}{2}\ri)
\ln\lf(1-e^{2\pi i(kT+lU+bC)}\ri)\ri\}\ .}
\eea
It can be checked that this expression stays finite for the
two limits $T\ra U$
and $C\ra0$. In the first case the state with $k=1,l=-1$ has to be
extracted from the instanton sum to cancel the same logarithmic singularity
arising from the Siegel forms.  In the second case these are the
states $k=0=l, b=-1$ and $k=0=l, b=-2$ which cause
the logarithmic singularity $-168/5\ln C$, which is cancelled by
a same term  from the Siegel forms. Moreover, it is not
too difficult to show that \req{final} becomes \req{final0} 
for $C\ra 0$. 
Again, for the large $T$--limit we obtain:

\be\label{LARGE}
\si(T,U,C)\lra -4\pi\im(T)\ \ \ ,\ \ \ T\lra i\infty\ .
\ee

The one--loop threshold correction $\triangle_a$ to the physical, i.e.
effective gauge couplings \req{triangle}, must be a regular and duality 
invariant quantity. This can also be inferred from the 
world--sheet torus integral. 
The holomorphic functions $\Psi_a^{16/5}$ in \req{dgl} have weight
$+120$ and $-120$, respectively. With the corresponding
$b_a$ from \req{beta} the left hand side of \req{dgl} 
becomes a duality invariant expression up to shifts caused
e.g. by $T\leftrightarrow U$ and $C\ra -C$.
Comparing the left hand side of \req{dgl} with \req{cindy} tells us, 
that the transformation behaviour of $\si$ is the same as 
that of $G_{N=2}^{(1)}$. 
On the other hand,
when we look at equations \req{triangle} and \req{ddd} we conclude that
$-iS+i\ov S+\fc{1}{8\pi^2} G^{(1)}_{N=2}=inv.$. This is precisely 
the object entering the K\"ahler potential
[cf. after eq. \req{mixing}] or eq. \req{finalgross}. 
We therefore write

\be
\fc{-iS_{\rm inv}+i\ov S_{\rm inv}}{2}:=\fc{-iS+i\ov S}{2}+\fc{1}{16\pi^2}
\si(T,U,C)=inv.
\ee
with: 

\be
-iS_{\rm inv}:=-iS+
\lf[\fc{2}{5}(\p_T\p_U-\fc{1}{4}\p^2_C)h^{(1)}(T,U,C)
\ri]-\fc{1}{5\pi^2}\ln\lf(\fc{\Delta_{35}^2}{\Delta_{10}^{22}}\ 
\chi_{12}^{25/2}\ri)^{1/4}\ .
\ee
Whereas the quantities $S$ and $\si(T,U,C)$ are shifted 
under $Z_2^{T\leftrightarrow U}$ and $C\ra e^{\pi i} C, C\ra e^{2\pi i}C$,
the field $S_{\rm inv.}$ is completely invariant  and it is
that combination which enters the physical gauge 
couplings \req{running}. 

Let us now turn to the GS--correction \req{mixing}. With \req{prep}
it becomes

\bea
\ds{G^{(1)}_{N=2}}&=&\ds{-\fc{8}{Y}
\pi\lf(\fc{2}{3}U_2^3+\fc{80}{3}C_2^3-14C_2^2U_2-12C_2^2T_2+const.\ri)+}\nnn
&+&\ds{\fc{4}{Y\pi}\re\lf\{\sum_{(k,l,b)>0}c_1\lf(kl-\fc{1}{4}b^2\ri)
\Pc\lf[e^{2\pi i(kT+lU+bC)}\ri] \ri\}\ ,}
\eea
with:

\be
\Pc\lf(e^{2\pi i x}\ri)=\fc{1}{2\pi}Li_3\lf(e^{2\pi i x}\ri)+\im(x)
Li_2\lf(e^{2\pi i x}\ri)\ .
\ee

We give two plots, one\footnote{The `instanton contributions' given by
the last term of \req{final} are taken into account up
to a certain order. This is necessary, e.g. in order to 
obtain the finite value for $C\ra 0$.} for $\sigma(T,U,C)$
and a second for $G^{(1)}_{N=2}(T,U,C)$:

\vspace{-0.5cm}
\vspace{1cm}
\epsfbox[-80 0 500 210]{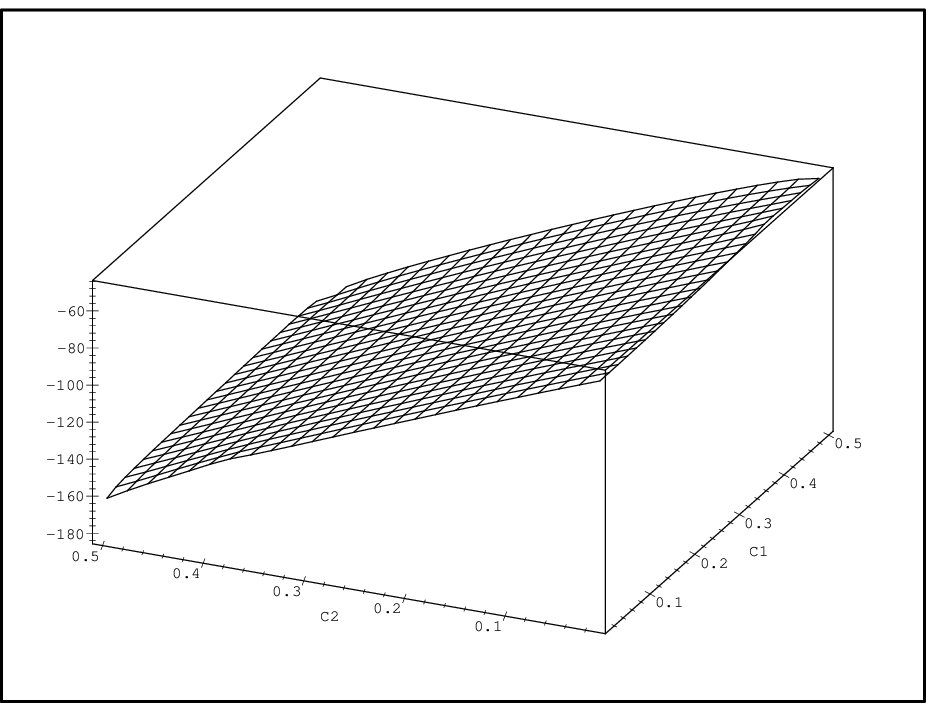}
\bc
{\small \em Fig.1 -- Dependence of the universal one--loop correction 
$\si$ on the Wilson line modulus $C$ for fixed $T=4i,\ U=i$.}
\ec

\vspace{0.0cm}
\epsfbox[-80 0 500 210]{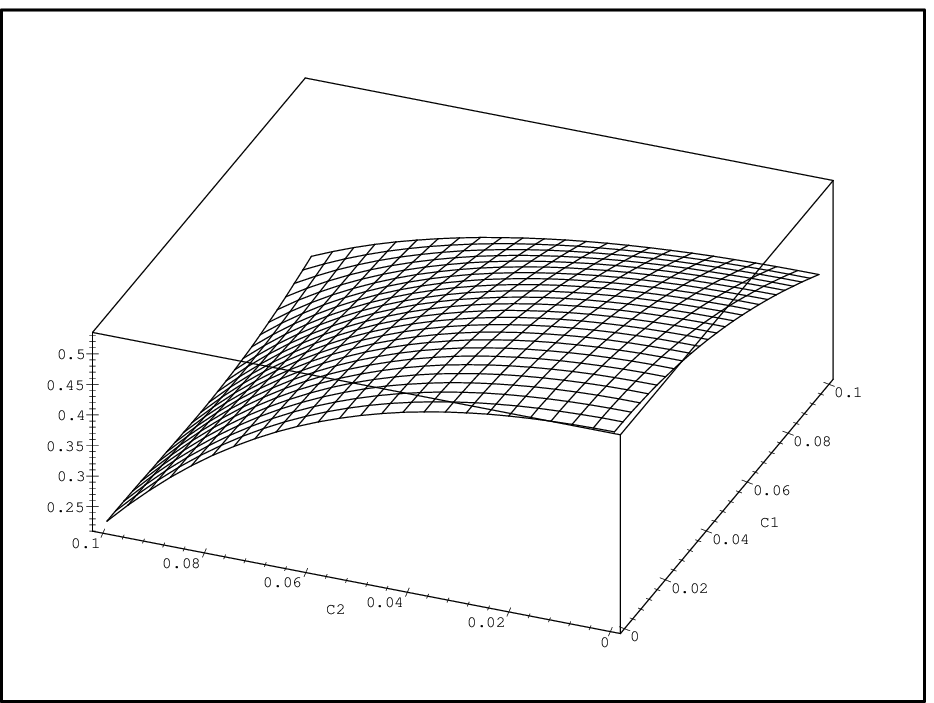}
\bc
{\small \em Fig.2 -- Dependence of the GS--correction
$G^{(1)}_{N=2}$ on the Wilson line modulus $C$ for fixed $T=4i,\ U=i$.}
\ec

\sect{Towards stronger coupling}

Let us now discuss the results of the previous threshold calculation from a
more general point of view. 
First, we summarize \req{triangle} and \req{ddd} in the holomorphic 
gauge kinetic function ($k_a=1$)

\be\label{ff}
f_a(S,T,U,C)=-iS+\fc{2}{5}(\p_T\p_U-\fc{1}{4}\p_C^2)h^{(1)}-
\fc{1}{5\pi^2}\ln\lf(\fc{\Delta_{35}^2}{\Delta_{10}^{22}}
\chi_{12}^{25/2}\ri)^{1/4}+\fc{b_a^{N=2}}{8\pi^2}\Delta^X_{holom.}\ .
\ee
Here $\Delta_{holom.}$ is just the holomorphic part
of \req{d1} and \req{d2}, for $X=I,II$ and $G_a=SO(12), E_8'$, respectively,
i.e.:

\be
\Delta^X_{holom.}=\lf\{\ba{c}
-\fc{2}{5}\ln \fc{\chi_{10}^{1/2}(\Om)}{\chi_{12}^{5/12}(\Om)}
-\fc{1}{12}\ln \chi_{12}(\Om)\ ,\ X=I\\[5mm]
-\fc{1}{12} \ln \chi_{12}(\Om)\ ,\ X=II \ea \ri.
\ee
Using identities of the previous section allows us to write alternatively

\be\label{f1}
f_a(S,T,U,C)=-i\tilde S-\fc{1}{5\pi^2}\ln\Psi_a(T,U,C)\ ,
\ee
with $\Psi_a$ defined in \req{psis} and the pseudo--invariant dilaton 

\be\label{f2}
-i\tilde S=-iS+\fc{2}{5}(\p_T\p_U-\fc{1}{4}\p^2_C)h^{(1)}(T,U,C)\ ,
\ee
which is invariant under the perturbative duality group up
to shifts \cite{dewit}. 
Before we proceed, let us introduce a different normalization of the dilaton
field $S$ in \req{S} and the $f$ functions \req{ff} by
$S\ra 4\pi S$ and $f\ra 4\pi f$. Thus $S$ becomes:

\be\label{newS}
S=\fc{\th_a}{2\pi}+i\fc{4\pi}{g_a^2}\ .
\ee
This form of $S$ is more convenient for the  study of duality 
symmetries like e.g.  
$S\leftrightarrow T$--exchange symmetry.

We want to investigate how much basic information
is encoded in the results given in \req{ff}. 
For this
let us take a step back and consider the whole situation from a field
theoretic point of view. There we know that the existence of nontrivial
threshold corrections can be deduced from the explicit anomaly
cancelation mechanism in the low energy $d=10$ field theory. For completeness
we repeat  this line of arguments. The anomaly cancelation mechanism 
discovered by Green and Schwarz requires additional terms in the action of
the low energy $d=10$ field theory: there are Chern--Simons terms
appearing in the definition in the 3-index field strength $H$ as well as
specific one loop counter terms \cite{GS}. 
It is these counterterms that are of
special interest for us, e.g.

\be\label{GSterm}
\epsilon^{KLMNOPQRST} B_{KL} Tr F^2_{MNOP} Tr F^2_{QRST} 
\ee    
where upper case latin indices  $K,L= 0\ldots9$ denote the
components of spacetime, $B$ is the two index antisymmetric tensor
and $F$ is the field strength of the Yang--Mills interactions. We are
interested in the compactification of this $d=10$ theory to a $d=4$ theory
with $N=1$ supersymmetry. Curvature terms $Tr R^2$ as well as field
strengths $Tr F^2$ will have nontrivial vacuum expectation values in
the extra six dimensions, fulfilling a consistency condition in order for
the three-form field strength $H$ with 
\be\label{dH}
dH = Tr F^2 - Tr R^2
\ee
to be well defined. Let us assume now that $Tr F^2_{klmn}$ is nonzero.
Lower case latin indices will represent the components of the
compactified 6 dimensions while greek indices will denote the
uncompactified four dimensions. The Green-Schwarz term given above 
\req{GSterm} will then reduce to

\be\label{axion}
\epsilon^{mn}B_{mn}\epsilon^{\mu\nu\rho\sigma}
Tr F_{\mu\nu} F_{\rho\sigma} 
\ee
in the $d=4$ theory. The gauge kinetic terms in that theory will
be given by the usual term $f(\phi)W^\alpha W_\alpha$ where at tree
level $f=-iS$ with $S$ being the dilaton superfield. 
An explicit inspection of the
$d=4$ action in components tells us that $\epsilon^{mn}B_{mn}$ is the
pseudoscalar axion that belongs to the $T$-superfield
\cite{Switten,din,DIN}. Upon 
supersymmetrization the term in \req{axion} will then lead to
a one loop correction to the holomorphic $f$--function that is
proportional to $T$ and its coefficient being fixed by the anomaly
\cite{DIN,in86}.
This is, of course, nothing else than a threshold correction. In the
simple case of the standard embedding leading to a $d=4$ theory
with gauge group $E_6 \times E_8$ one obtains \cite{in86}

\be\label{f6f8}
f_{E_6} = -iS -\epsilon iT\ \ \ \   {\rm and}\ \ \ \  f_{E_8}'= -iS + 
\epsilon iT
\ee
respectively, where $\epsilon$ is a constant fixed by the anomaly.
This should then be compared to the results given in eq. \req{ff}
Of course, the results in \req{f6f8} are obtained in the field theory
limit, i.e. the large radius limit of string theory and should therefore
correspond to a threshold calculation in the large $T$--limit.  

We now consider the decompactification limit $T\lra i\infty$ of \req{ff}.
Using the limits

\bea
\ds{\fc{\Delta_{35}^2}{\Delta^7_{10}}}&\lra&\ds{e^{2\pi i(-3T)}}\nnn
\ds{\Delta_{10}}&\lra&\ds{e^{2\pi iT}}\nnn
\ds{\chi_{12}}&\lra&\ds{e^{2\pi i T}\ ,}
\eea
we obtain from \req{f1} and \req{f2} with the changes made before eq.
\req{newS} 

\bea\label{lastlimits}
\ds{f_{SO(12)}} &\stackrel{T\ra i \infty}{\lra}& \ds{ -iS-6iT
=-iS-i\lf(\fc{b_{E_7}^{N=2}}{12}-1\ri)T\ .}\nnn
\ds{f_{E_8'}}   &\stackrel{T\ra i \infty}{\lra}& \ds{ -iS+6iT
=-iS-i\lf(\fc{b_{E_8'}^{N=2}}{12}-1\ri)T}
\eea
Note, how in the second limit the $E_7$ $\beta$--function appears.
In the limit $T\ra i\infty$ all $W$--bosons with masses $m^2\sim|C|^2/\im(T)$ 
become light and all states can be arranged in $E_7$ gauge multiplets.
To make contact with \req{f6f8} we have to clarify how 
our N=2 results \req{lastlimits} can be taken over
to the N=1 theory. The moduli dependence of all our results
arises from an N=2 subsector of our N=1 theory and therefore the 
$\beta$--functions are those of the N=2 theory. 
As a consequence of \req{splitting} they can be related to the 
N=1 $d=4$ anomaly coefficients

\bea
\ds{b_{E_8'}^{N=2}}&=&\ds{2\al^{N=1}_{E_8'}=-60}\nnn
\ds{b_{E_7}^{N=2}}&=&\ds{2\al^{N=1}_{E_6}=84\ ,}
\eea
if e.g. the N=2 subsector of our N=1 $d=4$ theory represents a standard
$Z_2$ orbifold and $f_{SO(12)}$ becomes $f_{E_6}$.

We have thus seen that the results of section 4 correspond exactly to
the anomaly cancelation terms in the large $T$--limit \req{f6f8}, 
and are thus to a
large extent determined by the mechanism of anomaly cancelation. This  result
gains further importance because of the fact that there exists 
a nonrenormalization theorem for the holomorphic $f$--function. No 
further correction do appear at higher loops, i.e. the above results
give the full perturbative result to all loops. This can be verified
by an inspection of the symmetries of the theory \cite{N} and general theorems
of N=1 supersymmetry \cite{SV}.

Let us now summarize what we have learned up to now. We have seen

\begin{itemize}

\item that the holomorphic threshold corrections in the large $T$--limit
are just a reflection of the anomaly in ten dimensions. 

\item that both the gauge dependent ($\Delta$) and the 
gauge independent threshold corrections ($\sigma$) are important for the
holomorphic $f$-function. Although $\sigma$ is not of great importance for the
question of unification it is crucial for the comparison of the
coefficients in the large--$T$ limit: $\Delta$ alone would lead to the
wrong result.

\item that the actual result of \req{f6f8} reveals a surprise that
was not appreciated enough when it was first obtained.
The limit of large $T$ will always lead to a situation where one of
the gauge groups becomes strongly coupled, independent of the size
of the original string coupling. The other coupling will become small
at the same time, a situation that has meanwhile been seen in various
other considerations \cite{DMW,fernando}
 that are, of course, based on the same argument
as the above, namely anomaly cancelation. This might be a first hint for
a conjecture that the results of the perturbative theory might
carry over to the strong coupling regime.
\end{itemize}

From the practical point of view it would be interesting to know how
well the large $T$--limit approximates the exact results. Although this
is model dependent we find in many cases that this approximation
holds even for smaller $T$--values. Let us just consider a few examples.
The gauge dependent thresholds $\Delta$ (see \req{logT}) are
proportional to $\ln\eta(T)$ and it turns out that this function
can be approximated very well by the linear function 
$i\pi T/12$. Even for values as small as $T=2i$ the linear
approximation is better than $10^{-3}$. In the gauge group--independent 
corrections $\sigma$ (see \req{final0}) there is a term
$\ln j(T)$ that is relevant in the large--$T$ limit. Again a linear
approximation is very accurate for values of $T>2i$.
Thus the results obtained in the large--$T$ limit
hold even for rather small values of $T$. In some cases one might
even deduce the full threshold corrections using symmetry
arguments like $T$--duality $SL(2,Z)$ 
 from the results obtained in the large--$T$ limit.
However, this mechanism works only in the simplest cases
\cite{DKL2}. Already for the examples considered in \cite{ms1} it 
does not lead to an unique answer.

As a side remark, let us add some comments about the question
of unification. We have seen in the previous sections that 
 with  vacuum expectation values of
the moduli fields $T$, $U$ and 
the Wilson line $C$ not too far away from the string scale,
the unification of the coupling constants can be achieved. This is in
contrast to the case without a Wilson line, where one has to choose
rather large values of $T$ and/or $U$ to achieve unification \cite{IL}. 
In this case it could be
argued (see e.g. \cite{BanksDine}) 
that this leads  in general  to strong
string coupling since the coupling constants evolve very fast
between the compactification scale and the
string scale. But there exist models where such a fast evolution
can be avoided \cite{AKS}.
Of course, we do not understand why $S$ should be large in the first
place, so we maybe would prefer stronger coupling for aestethic
reasons. At the moment, however, we can conclude that the
requirement of unification does not necessarily lead to strong
coupling. The problem with the strong coupling regime is the
the lack of ability to do 
reliable calculations and the study of unification 
might have to rely on wishful thinking.

To improve this situation
let us now investigate to what extent our exact calculations in the weak 
coupling regime could extend to the region of strong coupling. Our main
result is the fact that because of the anomaly and holomorphicity the
complete $f$-function is:

\be\label{fcomplete}
f_{\rm complete} = f_{\rm tree} + f_{\rm one-loop} + f_{\rm nonpert.},
\ee
where $f_{\rm tree}$ and $f_{\rm one-loop}$ are known in perturbation theory
[see e.g. \req{ff}].

Of course, we cannot say very much about the last term, although it
might be important for the question of the size of the coupling 
constant \cite{Lalak}.
The first term is given by the anomaly and we believe that this should
also be relevant in the region of stronger coupling. The strong coupling
limit of the $E_8 \times E_8$ heterotic string is believed to be 
the M--theory orbifold $S_1/Z_2$ as discussed by Horava and Witten
\cite{HoravaW}.
The eleventh dimension is an interval whose length $\rho$ becomes large
in the strong coupling limit. The gauge fields live on the ten--dimensional
boundaries of this interval, one $E_8$ on each side. Witten has 
investigated the question of unification of gauge and gravitational
couplings in the framework of this theory. He pointed out that 
 one could, in principle, push the Planck mass
to arbitrarily large values (by making $\rho$ large)
 and thus adjust the gravitational coupling to
any desired value while keeping the gauge coupling fixed
\cite{newwitten}. However,
in many cases, there is an obstruction to this adjustment once
quantum effects are incorporated. To see that we can consider a 
$d=4$, $N=1$ theory with gauge group $E_6 \times E_8$ obtained via
compactification on a Calabi-Yau manifold ($X$). Viewed from the eleven
dimensional theory, the compactification is not a direct product of
$X$ and $S_1/Z_2$ as one might have naively guessed. The reason for it is a
nontrivial vacuum expectation value of the four index field strength
$G_{klmn}$ of the eleven dimensional theory, that appears because of
a consistency condition similar to \req{dH}. In the ten dimensional
theory we can satisfy \req{dH} by the standard embedding

\be
Tr F_6^2 = Tr R^2
\ee
with $Tr F_8^2=0$. In M--theory there is, however, a contribution 
$Tr F_i^2 - {1\over 2}Tr R^2$  at each boundary. Via supersymmetry this then
induces nontrivial values for the $G$--field at the boundaries.
Witten has solved the conditions for unbroken supersymmetry in a
linearized approximation to first order \cite{newwitten}. 
This leads to a solution
where the size of the Calabi-Yau manifold varies with the 
eleven dimensional coordinate. The limiting case where the size of
$X$ vanishes at the $E_8$ boundary then leads to an upper limit on
the possible size of the eleven-dimensional interval, which in turn
leads to an upper limit on the Planck mass. In fact the radii at the
different boundaries are given in the linearized approximation by

\be\label{R6R8}
R_8\sim S- \epsilon T\ \ \ \ \ {\rm and} \ \ \ \  R_6\sim 
S+\epsilon T
\ee
where the coefficient $\epsilon$ is the same as in \req{f6f8}.
The size of $X$ vanishes on the $E_8$ side at the same point at which
$f_8$ vanishes in the calculation done in the weak coupling
regime (where the coupling of $E_8$ will become large).
The similarity of \req{f6f8} and \req{R6R8} is, of course, not
an accident and should be no surprise. 
In both cases the reason for the result comes from the
mechanism of the cancelation of anomalies, and our large-$T$  limit in
the weakly coupled heterotic string \cite{in86} coincides with 
Witten's linearized approximation in the strongly coupled
region\footnote{The similarity between 
these calculations has also been 
observed in \cite{BanksDine}}\cite{newwitten}.
 This shows that
the result of \req{f6f8} is very robust and carries over to the
strong coupling regime, the reason being the holomorphicity of $f$
and the relation to the anomaly. \req{ff} allows us to go beyond
the large $T$ limit of \req{f6f8} and the linearized approximation
of \req{R6R8}.
In the strongly coupled theory it
is still valid, but has a different (now geometrical) interpretation.
Moving the Planck scale to a large value in M-theory is nothing else
than a threshold calculation in the weakly coupled heterotic
theory. At large coupling we should, however, be aware of the fact
that nonperturbative effects like $\exp(-S)$ might become important.
There is the hope that such terms could be determined with the
help of string dualities 
but at the moment we have to rely on simplified assumptions. Such
corrections might be important for the relation of $S$ to the
fundamental string coupling constant and might be the reason why
$g_{\rm string}$ is small even if $S\sim 1$ \cite{Lalak}.

So far we have only concentrated on the discussion 
of $T$ (the case of $U$ being equivalent) and not yet the Wilson line.
  In the last section, we have seen the
exact results for the thresholds in the perturbative theory
including the Wilson lines. Are these
results also valid in the limit of stronger coupling. This
question is not so easy to answer, because the relation of the
Wilson lines 
to the mechanism of anomaly cancelation is more complicated. 
Still they give a 
contribution to the holomorphic $f$--function.
We can have a look at the anomaly cancelation terms as in
\req{GSterm}, which contain the pseudoscalar axion that is related
to the $T$--multiplet as well as to the Wilson line. In the weakly 
coupled theory the kinetic terms are determined by the K\"ahler potential.
In general with a Wilson $A$:

\be 
K = -\log(S+\bar S) -3\log(T + \bar T -  2 A\bar A)\ . 
\ee
Thus the mixing of $T$ and
$A$ which determines the  scalar partner of the axion 
$\epsilon^{mn}B_{mn}$ is controlled by the K\"ahler potential.
As of now, we do not know any useful nonrenormalization theorems for
the K\"ahler potential in N=1 supersymmetric theories. Corrections
might appear at any loop level and thus the weak coupling 
results are not likely to carry over to the strong coupling regime.
Of course, the form of the $f$--function at one loop will still
remain the complete perturbative result. But there is no easy way to
deduce it from first principle, {\it it has to be computed in the framework
of string theory}. More exact information about the K\"ahler potential
might be obtained in theories with extended supersymmetry, where there
is a relation to the holomorphic prepotential. So the question
how the threshold results with Wilson line carry over to strong coupling
is still open. It is certainly worthwhile to have a closer look at this 
question in the future.

\sect{Conclusions}

The main technical achievement presented in this paper is a full
calculation of the gauge independent
thresholds in the presence of a continuous
Wilson line \req{final}. 
Equipped with these results we can
draw important conclusions for the following two questions:

\begin{itemize}

\item Is the (perturbative) heterotic string able to describe
the unification of gauge and gravitational coupling constants?

\item Does the perturbative result give us information about the
limit of strong coupling? 
\end{itemize}

The first question can be answered with yes, as we pointed out earlier 
already \cite{ns}. 
It is not necessary to go to strong coupling to achieve unification.
With moderate values of the $T$, $U$ and $C$ moduli unification can
be achieved. The role of the Wilson lines is very important. {\it It shows 
that a discussion of string unification without the consideration of such 
Wilson lines is not very meaningful}. Let us also stress that the
gauge group dependent thresholds $\Delta$ are of great importance for
unification. The universal terms $\sigma$ although quite large in
some cases are not as important in that respect. The question, why
the size of the string coupling constant is small compared to unity
is not yet understood in this scenario, but this question will
strongly depend on the size of nonperturbative effects.

The answer to the second question is positive as well. We have shown
that the perturbative calculation for the $f$-function is even reliable
in the regime of stronger coupling. The reason for that is the
nonrenormalization of $f$ beyond one loop due to its holomorphic
structure as well as the close connection of the thresholds to
anomaly cancelation. Here both $\Delta$ and $\sigma$ play an
important role. The calculation of the thresholds at weak coupling
in the large $T$ limit
coincide with those of strongly coupled M-theory in the linearized
approximation (although with a different geometrical interpretation).
{\it The full result \req{ff} allows us to go beyond the linearized
approximation of M--theory}. 
The separate role of $T$, $U$ and $C$ depends, however,
on the loop corrections to the K\"ahler potential that cannot be easily
controlled in theories with $N=1$ supersymmetry. Therefore the
dependence of $f$ on the Wilson line cannot be easily obtained from
field theoretic arguments and has to be computed in string theory.
It remains an open question how to obtain reliable results 
for the K\"ahler potential 
in the strong coupling limit of $N=1$ theories. 
Certainly extended supersymmetry could
give more restrictions. In addition, any information about possible
nonperturbative contribution to the $f$-function will be extremely
useful both from a theoretical and phenomenological point of view.
It might ultimately lead to a situation where the value of $S$ is fixed,
implying an understanding of the size of the string coupling.
It might also be decisive for an understanding of the dynamical
breakdown of supersymmetry.
Hopefully string theory dualities will eventually give some information 
in that direction.

\ \\ \\
{\em {\mbox{\boldmath $Acknowledgement:\ $}}}
We are very grateful to J.--P. Derendinger for valuable discussions
especially related to section two. Moreover we thank G.L. Cardoso, 
K. F\"orger, E. Kiritsis, J. Louis, P. Mayr, F. Quevedo 
and S. Theisen for discussions.

\appendix
\section*{Appendix}

This appendix is taken from \cite{ich}.

\sect{Siegel forms of weight $k$ and degree $g=2$}

{\em Siegel modular forms} are a natural generalization of elliptic modular
forms to the higher genus case.

\ \\
{\bf Definition 1:}\ A {\em Siegel modular form} $f(\Om)$ of degree $g$ and 
weight $k$
is defined by the following two conditions:

\begin{enumerate}
\item For every element $M \in Sp(2g,\Z)$, $f(\Om)$ satisfies:

$$f(M\Om)=\det (\Cc\Om+\Dc)^k f(\Om)\ ,\ k \in \Z\ ,$$
with $M \cong \lf(\ba{cc}
 \Ac&\Bc\\  \Cc&\Dc \ea \ri) \in Sp(2g,\Z)$\ ,\  i.e.: $\Ac^t\Cc=\Cc^t\Ac\ ,\ 
\Dc^t\Ac-\Bc^t\Cc={\bf 1}_g$ and $M\Om:=(\Ac\Om+\Bc)(\Cc\Om+\Dc)^{-1}\ .$
\item It is holomorphic for $\Om = \lf(\ba{cc}
 T&C\\  C&U \ea \ri) \in \Fc_g$, 
{\em Siegels fundamental region} for genus $g$. 
\end{enumerate}
Examples of them are the {\em Eisenstein series}:

\be
\Ec_k(\Om)=\sum_{\Cc,\Dc} \fc{1}{\det (\Cc\Om+\Dc)^k}\ ,
\label{eisen2}
\ee
where the summation runs over all inequivalent bottom rows of elements of 
$Sp(2g,\Z)$ with respect to 
left multiplication by unimodular integer matrices of 
degree $g$.

The graded ring of the modular forms is generated for $g=1$ by the two
Eisenstein series $E_4$ and $E_6$. Igusa has  proven an analogous result
for $g=2$  \cite{igusa}: The Eisenstein series of $g=2$: 
$\Ec_4,\Ec_6,\Ec_{10}$ and $\Ec_{12}$  are algebraically independent over 
$\CC$ and they generate 
the graded ring of the modular forms for $g=2$ and even weight.
As in the $g=1$ case one is interested in the cusp forms. They are 
constructed out of $\Ec_4,\Ec_6,\Ec_{10},\Ec_{12}$.
There are {\em two cusp forms} in the $g=2$ case, one of weight 10 and
the other of weight 12, respectively \cite{igusa}:

\bea
\chi_{10}(\Om)&=&\ds{c_1\lf[\Ec_4(\Om)\Ec_6(\Om)-\Ec_{10}(\Om)\ri]}\nnn 
\chi_{12}(\Om)&=&\ds{c_2\lf[441\Ec_4^3(\Om)+250\Ec_6^2(\Om)
-691\Ec_{12}(\Om)\ri]\ .}
\eea
with the numerical constants $c_1=-43867/(371\ 2^{12}3^55^2)$ and 
$c_2=77683/(2^{13}3^75^37^2337)$.
For arbitrary weight $k \in \Z$ one has in addition a cusp form $\chi_{35}$ of 
weight $35$.
Alternatively the graded ring of of modular forms is generated 
by $\Ec_4,\Ec_6$ and the three cusp forms $\chi_{10}, \chi_{12}, \chi_{35}$.

All the cusp forms introduced before may be also expressed in terms of the
genus--two theta functions. The most general $g=2$ theta--function is 
defined by:
\nwc{\bz}{\mbox{\boldmath $z$}}
\be
\vartheta\lf[ \ba{c} a_1 \atop b_1 \\a_2\atop b_2 
\ea \ri](\Om,\mbox{\boldmath $z$})=\sum_{n_1,n_2 \in \Z} 
e^{\pi i\mbox{\boldmath $(n+a)$}^t
\Om\mbox{\boldmath $(n+a)$}+2\pi i\mbox{\boldmath $(n+a)$}^t
\mbox{\boldmath $(z+b)$}}\ .
\ee
There are sixteen of them: ten of even characteristics and six of odd 
characteristics \cite{freitag}. 
The ten even characteristics are:

$$\lf[ \ba{c} a_1 \atop b_1 \\a_2\atop b_2 
\ea \ri]\in \h\lf\{ \ba{cccccccccc} 0&0&0&0&1&1&0&0&1&1\\
                                    0&1&0&1&0&0&0&1&0&1\\
                                    0&0&0&0&0&0&1&1&1&1\\
                                    0&0&1&1&0&1&0&0&0&1\ea\ri\}\ .$$
We can represent our cusp forms by $\vartheta$--characteristics:

\bea\label{chi10chi12}
\chi_{10}(\Om)&=&\ds{-\fc{1}{2^{14}}\prod_{even}\vartheta^2_a(\Om,0)}\nnn
\chi_{35}(\Om)&=&\ds{-\fc{i}{2^{39}5^3}\lf[\prod_{even}\vartheta_a(\Om,0)\ri]
\sum_{(a,b,c)\in\ azygous}\de_{abc}\ \vartheta_a^{20}(\Om,0)
\vartheta_b^{20}(\Om,0)\vartheta_b^{20}(\Om,0)\ .}
\eea
There are 60 {\em azygous} triples $(a,b,c)$ denoting 
a certain combination of $\vartheta$--characteristics and 
$\de_{abc}=\pm 1$ to ensure the correct behaviour under $Sp(4,\Z)$.

To obtain an expression for $\chi_{12}$ in terms of 
$\vartheta$--characteristics we first consider two theorems,
which allow one to represent the  
limit $B\ra 0$ of a modular form of genus 2 in a product of well--known 
genus one functions \cite{freitag}:

\ \\
{\bf Theorem 2:} If $F$ is a modular form of $g=2$ and weight $k=0\mod 2$,
then $F\lf(\ba{cc}
 T&0\\ 0&U \ea \ri)$ can be represented as isobar 
polynomial in the functions:

$$E_4(T)E_4(U)\ \ ,\ \ E_6(T)E_6(U)\ \ ,\ \ E_4^3 (T) E_6^2(U)+E_4^3(U)E_6^2(T)
\ .$$
\ \\
{\bf Theorem 3:} There are modular forms $F_k$ of $g=2$ with degree
$k=4,6,12$, respectively and:

\be
F_k\lf(\ba{cc}
 T&0\\ 0&U \ea \ri)=\lf\{ \ba{ll}E_4(T)E_4(U)\ ,&k=4\\
                  E_6(T)E_6(U)\ ,&k=6\\
                  E_4^3 (T) E_6^2(U)+E_4^3 (U) E_6^2(T)\ ,&k=12\ . \ea\ri.
\ee
The modular functions

\bea
\ds{F_4(\Om)}&=&\ds{\fc{1}{4}\sum_{even}\vartheta_a^8(\Om,0)}\nnn
\ds{F_6(\Om)}&=&\ds{\fc{1}{4}\sum_{(a,b,c) \in syzygous}\de_{abc}\ 
\vartheta_a^4(\Om,0)\vartheta_b^4(\Om,0)\vartheta_b^4(\Om,0)}\nnn 
\ds{F_{12}(\Om)}&=&\ds{-\fc{81}{44}\ \h
\sum_{even} \vartheta_a^{24}(\Om,0)+\fc{11}{2}F^3_4(\Om)+\fc{2}{11}F_6^2(\Om)}
\label{c12}
\eea
are constructed to have this behavior for $B=0$.
Again there are 60 {\em syzygous} triples $(a,b,c)$ denoting 
a certain combination of $\vartheta$--characteristics and 
$\de_{abc}=\pm 1$. The modular function $F_{35}$ consists of 60 
{\em azygous} combinations.
All these functions can also be expressed by the $g=2$ Eisenstein functions 
\req{eisen2}. 
In fact we have $F_4(\Om)=E_4(\Om)\ ,\ F_6(\Om)=E_6(\Om)$. 
With a $F_{10} \sim \chi_{10}$ every $g=2$ modular form of even weight
can be represented as isobar polynomial in $F_4,F_6,F_{10}$ and $F_{12}$
\cite{freitag}.
Finally with \req{c12} we obtain an expression for $\chi_{12}$
in the sense of \req{chi10chi12}:

\be
\chi_{12}(\Om)=\fc{1}{1728^2}\lf[F_4^3(\Om)+F_6^2(\Om)-F_{12}(\Om)\ri]
\ee

The factorization properties in the theorems 2 and 3 originate
from the factorization properties of $\vartheta$--characteristics \cite{igusa}:

\be 
\vartheta\lf[ \ba{c} a_1 \atop b_1 \\a_2\atop b_2 \ea \ri]\lf(\ba{cc}
 T&C\\ C&U \ea \ri)=\sum_{n=0}^\infty \fc{2^{2n}}{(2n)!} \fc{d^n}{dT^n} 
\theta\lf[a_1 \atop b_1\ri](T)\fc{d^n}{dU^n} \theta\lf[a_2 \atop b_2\ri]
(U)C^{2n}\ .
\ee


\end{document}